# Pseudosymmetry in Tetragonal Perovskite SrIrO$_3$ Synthesized under High Pressure

Haozhe Wang[1], Alberto de la Torre[2], Joseph T. Race[3], Qiaochu Wang[2], Jacob P. C. Ruff[4], Patrick M. Woodward[3], Kemp W. Plumb[2], David Walker[5], Weiwei Xie[1]*

1. Department of Chemistry, Michigan State University, East Lansing, MI, 48824
2. Department of Physics, Brown University, Providence, RI, 02912
3. Department of Chemistry and Biochemistry, The Ohio State University, Columbus, OH, 43210
4. Cornell High Energy Synchrotron Source, Cornell University, Ithaca, NY, 14853
5. Lamont Doherty Earth Observatory, Columbia University, Palisades, NY, 10964

* Email: xieweiwe@msu.edu

*Abstract*

In this study, we report a tetragonal perovskite structure of SrIrO$_3$ (*P4/mmm*, $a = 3.9362(9)$ Å, $c = 7.880(3)$ Å) synthesized at 6 GPa and 1400 °C, employing the ambient pressure monoclinic SrIrO$_3$ with distorted *6H* structure as a precursor. The crystal structure of tetragonal SrIrO$_3$ was evaluated on the basis of single crystal and powder X-ray diffraction. A cubic indexing was observed attributed to overlooked superlattice reflections. Weak fractional peaks in the H and K dimensions suggest possible structure modulation by oxygen defects. Magnetization study reveals weak paramagnetic behavior down to 2 K, indicative of the interplay between spin-orbit coupling, electron correlations, and crystal electric field. Additionally, measurements of electrical resistivity display metallic behavior with an upturn at about 54 K, ascribed to weak electron localization and possible structural defects.

## Introduction

Iridates is an ideal system for investigating the intriguing quantum physics arising from the significant spin-orbit coupling (SOC). The exotic electronic quantum phenomena primarily arise from the interplay between the robust SOC and other fundamental interactions which have comparable energy scales, including crystal electric field and electron correlations, leading to competition between them. The SOC is typically negligible in the 3$d$ transition metal systems. However, in the case of 5$d$ transition metal compounds, such as the iridates, the SOC plays a crucial role in the observed physics behaviors, for example, superconductivity[1], Kitaev quantum spin liquid[2], Weyl semimetal behavior[3,4], quantum criticality[5,6], and quantum Hall effects[7]. Among iridate compounds, the Ruddlesden–Popper phase $Sr_{n+1}Ir_nO_{3n+1}$, specifically, $Sr_2IrO_4$ ($n = 1$) and $Sr_3Ir_2O_7$ ($n = 2$), have been intensively studied for novel physics with strong SOC on Ir sites.[8-12] In $Sr_2IrO_4$ and $Sr_3Ir_2O_7$, [$IrO_6$] octahedra arrange along the $ab$-plane forming the two-dimensional patterns. Conversely, owing to the enhanced interlayer hopping induced by the three-dimensional continuous arrangement of [$IrO_6$] octahedra, in comparison with $Sr_2IrO_4$ and $Sr_3Ir_2O_7$ which are insulators with canted antiferromagnetic ordering, $SrIrO_3$ ($n = \infty$) exhibits a paramagnetic semimetal ground state.[13,14] Thus, $SrIrO_3$ is positioned in proximity to a multi-phase boundary where the transitions of metal-insulator and magnetic ground states occur. These transitions are governed by the combination of multiple interactions, resulting in a complex interplay of competing effects.[13]

Instead of the bulk, most of current $SrIrO_3$ studies focus on constructing heterostructure thin films based on $SrIrO_3$ and other $ABO_3$ perovskite oxides (for example, $CaMnO_3$) using molecular beam epitaxy along particular crystallographic orientations, such as the [111] direction and the [001] direction.[15-18] The component layers require thickness at unit cell scale and lattice quality at atomic level, which sets a big challenge in sample preparation. On the other hand, the high pressure and high temperature synthesis provides a new approach to novel materials.[19] Recently, the non-centrosymmetric $Sr_2IrO_4$ was first discovered at 6 GPa and 1400 °C,[20] which further inspires us to investigate $SrIrO_3$ under high pressure and high temperature conditions.

Herein, we report the synthesis, structural characterization, magnetic and electrical resistivity study on the tetragonal $SrIrO_3$ synthesized at 6 GPa and 1400 °C for 3 hours (serial number TT-1446). Employing single crystal and powder X-ray diffraction (XRD), the tetragonal symmetry was carefully examined, excluding the possibility of pseudo cubic symmetry observed. Moreover, weak fractional peaks with the tetragonal space group in the H and K dimensions in the reciprocal space suggest possible structure modulation by oxygen defects. The magnetization study of tetragonal $SrIrO_3$ shows weak paramagnetic behavior down to 2 K. Furthermore, our electrical resistivity measurement demonstrates metallic property with a resistivity upturn at ~54 K due to weak electron localization and possible structural defects in the presence of SOC.

## Experimental Section

**High-Pressure Synthesis.** The high-pressure synthesis was conducted using a Walker-type[21] multi-anvil apparatus (MA) at Lamont-Doherty Earth Observatory. The starting material was as-synthesized ambient pressure $SrIrO_3$ phase, which was prepared by thoroughly mixing the materials $SrCO_3$ and Ir, and subsequently heating them to 900 °C for 12 hrs, then regrinding and reannealing at 1000 °C for 72 hrs.[22] The sample was kept at 120 °C overnight to remove the moisture before high-pressure experiments. The sample was then loaded in a platinum capsule inside an $Al_2O_3$ crucible that was inserted into a Ceramacast 646 octahedral pressure medium lined on the inside with a $LaCrO_3$ heater and kept at 6 GPa and 1400 °C for 3 hours before quenching to room temperature and then decompressed to ambient pressure overnight (serial number TT-1446). The pressure applied in the press has been calibrated using the phase transitions of bismuth at room temperature. To ensure accurate temperature readings, thermocouples are positioned as close as possible to the sample for direct temperature measurement.

**Chemical Composition Determination.** The recovered high-pressure product was examined for purity and homogeneity using a Zeiss Sigma field emission scanning electron microscope (SEM) with Oxford INCA PentalFETx3 energy-dispersive spectroscopy (EDS) system (model 8100). An accelerating voltage of 20 kV was employed for imaging and analysis.

**Phase Analysis by Powder X-ray Diffraction.** The phase identity and purity were examined using a Bruker Davinci powder X-ray diffractometer with Cu $K_\alpha$ radiation ($\lambda = 1.5406$ Å). Room temperature measurements were carefully performed with a step size of 0.010° at a scan speed of 5.00 sec/step over a Bragg angle ($2\theta$) range of 5–120°.

**Home Laboratory Single Crystal X-ray Diffraction.** The room temperature crystal structure was determined using a Bruker D8 Quest Eco single crystal X-ray diffractometer, equipped with Mo $K_\alpha$ radiation ($\lambda = 0.7107$ Å) with an $\omega$ of 2.0° per scan and an exposure time of 10 s per frame. A SHELXTL package with the direct methods and full-matrix least-squares on the $F^2$ model was used to determine the crystal structure.[23,24]

**Synchrotron Single Crystal X-ray Diffraction.** The experiments were carried out on the $QM^2$ beamline at the Cornell High Energy Synchrotron Source (CHESS). The sample was affixed to a Kapton mount using GE varnish and all data was collected at room temperature. We used an X-ray energy of 18 keV (0.6888 Å wavelength) and data was collected with a Pilatus 6M photon counting detector operating at a 10 Hz framerate while continuously rotating the sample angle (psi) through 360 degrees (gives one image every 0.1 degree). We combined scans taken with the sample angle chi angles of 0, 90, and 100 degrees in order to fill in gaps in the detector and reduced the data using a cubic unit cell, lattice parameters 3.937 Å.

**Physical Properties Measurement.** Temperature and field-dependent magnetization and electrical resistivity measurements were performed with a Quantum Design DynaCool physical property measurement system (PPMS) at a temperature range of 1.8–300 K and applied fields up to 9 T using 11.0 mg sample in total. Electrical resistivity measurements were conducted with a four-probe method using platinum wires on a polycrystalline sample in the dimensions of 1.0 mm × 0.8 mm × 0.5 mm.

## Results and Discussion

**Crystal Structure Determination and Phase Analysis.** The ambient pressure SrIrO$_3$ (m$C$-SrIrO$_3$) crystallizes in a monoclinic distortion of the hexagonal BaTiO$_3$-type structure. Under pressure, an orthorhombic perovskite SrIrO$_3$ (o$P$-SrIrO$_3$) in the GdFeO$_3$-type structure was reported at above 20 kbar at 1650 °C and above 50 kbar at 700 °C.[25] Here we performed our high pressure synthesis at 6 GPa and 1400 °C. Prior to pressurization, the precursor phase, as-synthesized m$C$-SrIrO$_3$, was confirmed by powder XRD, with the pattern and Rietveld refinement presented in **Figure 1a**. Around 7% SiO$_2$ (by weight) was added as an impurity phase to account for peaks that could not be indexed to m$C$-SrIrO$_3$ (refer to **Tables S1–S4** for details). **Figure 1b** shows the powder XRD pattern of our recovered product, where diffraction peak positions clearly illustrate the appearance of a new phase after high pressure and high temperature treatment.

To determine the crystal structure of the recovered product as well as initialize Rietveld refinements of the powder XRD pattern, single crystal XRD experiments were first completed at home laboratory. Limited by onsite X-ray diffraction data quality, two structure models were proposed. One is cubic perovskite structure ($a$ = 3.9403(6) Å) with the space group $Pm$-3$m$ (#221), denoted as c$P$-SrIrO$_3$, the other is tetragonal structure with $c$ lattice parameter twice of the others ($a$ = 3.9362(9) Å, $c$ = 7.880(3) Å, t$P$-SrIrO$_3$) in $P4/mmm$ (#123). The cubic structure model was slightly preferred. Crystallographic data and our single crystal XRD refinement details were summarized in **Tables S5–S8**. **Figure S1** provides regenerated reciprocal lattice planes from single crystal XRD based on these two structure models. Additionally, homogenous chemical element distribution of the recovered product was confirmed by scanning electron microscopy (SEM) and energy-dispersive spectroscopy (EDS) analysis on a single crystal sample (**Figures S2** and **S3**).

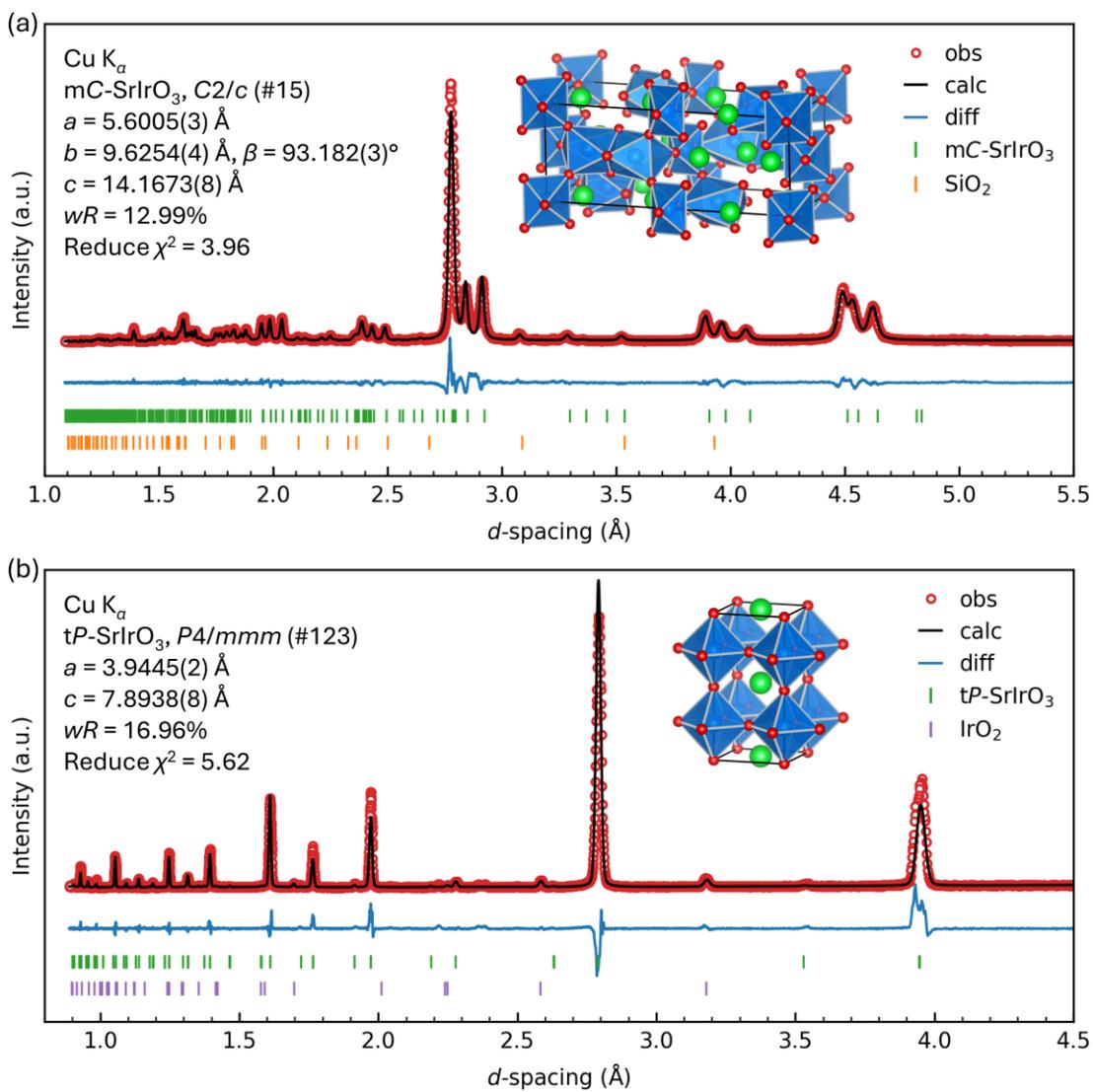

**Figure 1** Powder XRD pattern and Rietveld refinement of ambient pressure m$C$-SrIrO$_3$ and high pressure t$P$-SrIrO$_3$ synthesized at 6 GPa and 1400 °C. Bragg peak positions of each phase are represented by vertical tick marks. The crystal structures are also shown. Sr, green; Ir, blue; O, red.

c$P$-SrIrO$_3$ and t$P$-SrIrO$_3$ were then taken as starting parameters for the Rietveld refinements. Considering the inclusion of about 3% IrO$_2$ (by weight) as a secondary phase, both structure models yielded reasonable refinement parameters (details provided in **Tables S9–S12**). A minor difference was noted between these two models specifically within the $d$-spacing range of 0.9–1.5 Å. Unlike t$P$-SrIrO$_3$, c$P$-SrIrO$_3$ failed to address extra weak but clearly resolved diffraction peaks, as presented in **Figure 2**, suggesting a lower symmetry. The reported o$P$-SrIrO$_3$ (*Pbnm*, $a = 5.5909(1)$ Å, $b = 7.8821(1)$ Å, $c = 5.5617(1)$ Å)[22,26] model was also evaluated, however, the absence of diffraction peaks indicative of orthorhombic symmetry excludes its possibility as the optimized structure model (details shown in **Figure S4**, **Tables S13** and **S14**).

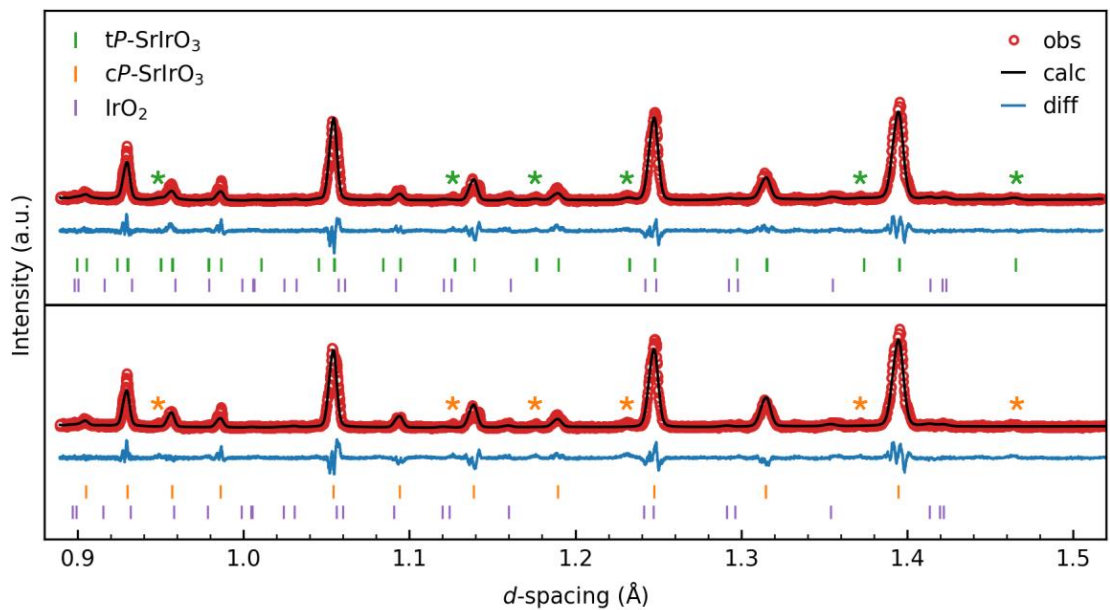

**Figure 2** Zoom-in powder XRD pattern of the recovered product and Rietveld refinement using t$P$-SrIrO$_3$ and c$P$-SrIrO$_3$ models. Bragg peak positions of each phase are represented by vertical tick marks. The crystal structures are also shown. Sr, green; Ir, blue; O, red. Stars indicate diffraction peaks for which the c$P$-SrIrO$_3$ model failed to account in comparison to t$P$-SrIrO$_3$.

To further assess the accuracy of the cubic indexing suggested by our in-house single crystal XRD analysis for the crystal structure of our recovered product, we carried out synchrotron single crystal XRD experiments at the Cornell High Energy Synchrotron Source (CHESS). For a better comparison, we reduced our data using cubic symmetry and the unit cell of c$P$-SrIrO$_3$ structure model. **Figure 3** provides the regenerated reciprocal lattice planes (HK-1), (HK-1.5), and (HK-2). The clearly resolved, relatively intense peaks on the (HK-1.5) plane corroborate the unit cell doubling along the $c$-axis, aligning with our t$P$-SrIrO$_3$ structure model. Moreover, weaker fractional diffraction peaks in the H and K dimensions suggest possible structure modulation due to oxygen defects in t$P$-SrIrO$_3$. Furthermore, the (HK-2) plane reveals distinct incommensurate peaks likely resulting from twinning in the single crystals.

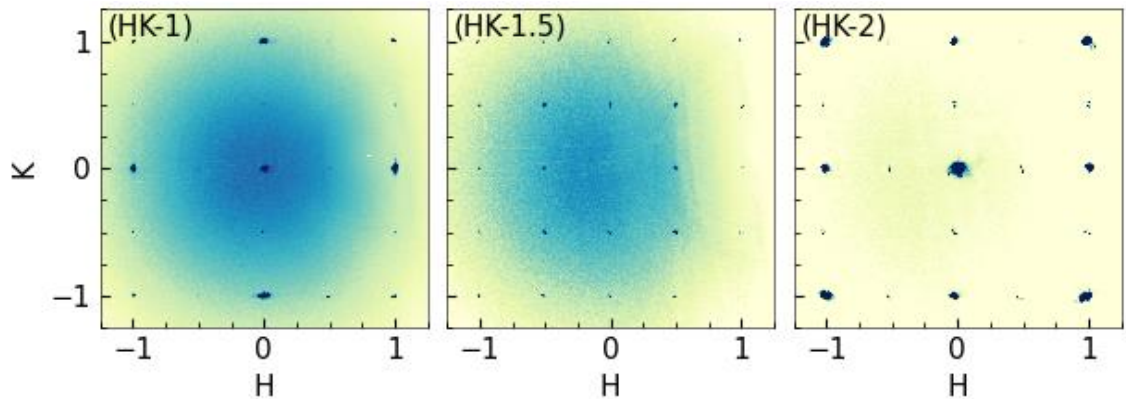

**Figure 3** Regenerated reciprocal lattice planes, (HK-1), (HK-1.5), and (HK-2) obtained from data reduction of our synchrotron single crystal XRD using cubic symmetry and the unit cell of c$P$-SrIrO$_3$.

Following the discussion, we identified the optimized structure model of our recovered product as t$P$-SrIrO$_3$ in space group $P4/mmm$ (#123) with the unit cell parameters $a = b = 3.9362(9)$ Å and $c = 7.880(3)$ Å. **Figure 4** presents a comparative analysis of the crystal structure and lattice parameters of c$P$-SrIrO$_3$, t$P$-SrIrO$_3$, and o$P$-SrIrO$_3$. Overall, t$P$-SrIrO$_3$ shows a ~4.0% volume compression per chemical formula compared to m$C$-SrIrO$_3$ under ambient pressure. The average Sr–O and Ir–O bond lengths measure 2.77(4) Å and 1.969(1) Å, respectively. The asymmetric Ir-O distances in octahedral IrO$_6$ indicate the existence of polar local distortions. The phase was quenched from high pressure high temperature synthesis and thus can be metastable at ambient conditions. The external pressure is a possible reason for the polar local distortion to be stabilized. Unlike common observations in ambient and high pressure SrIrO$_3$[22,26] and Sr$_2$IrO$_4$[20,27], the absence of [IrO$_6$] octahedra rotation, tilt, and oxygen site disorder indicate potential linear Ir–O–Ir exchange interactions in this t$P$-SrIrO$_3$.

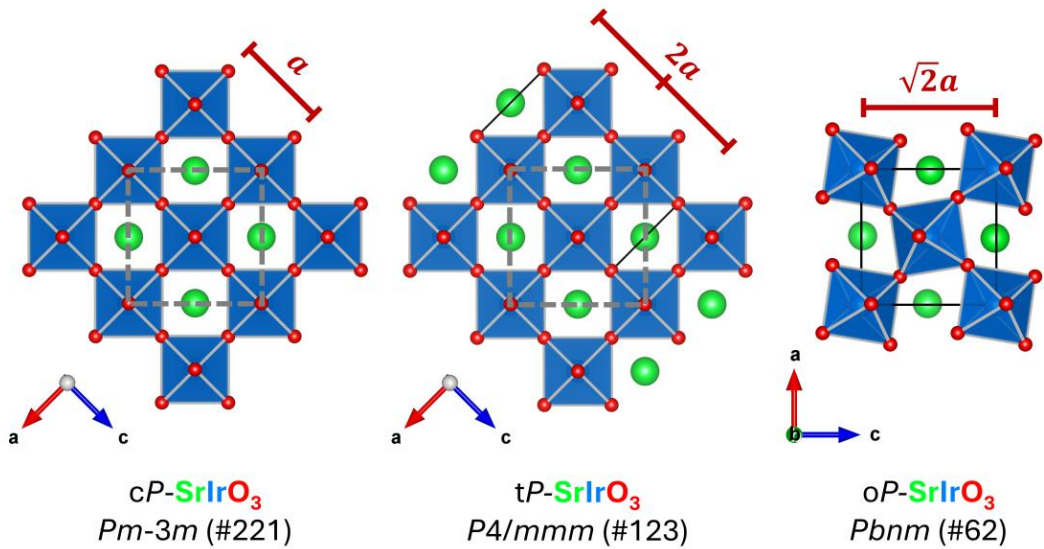

**Figure 4** Crystal structure and lattice parameter comparison of c*P*-SrIrO$_3$, t*P*-SrIrO$_3$, and o*P*-SrIrO$_3$. Sr, green; Ir, blue; O, red.

**Magnetization: Temperature-dependent Paramagnetism.** Similar to monoclinic SrIrO$_3$ and orthorhombic SrIrO$_3$, a paramagnetic susceptibility was observed in tetragonal SrIrO$_3$, as displayed in **Figure 5a**. There is no obvious split between zero-field cooled (ZFC) and field-cooled (FC) modes (See **Figure S5** for details). The initial examination of the data highlighted the small value of positive magnetic susceptibility and its noisy temperature dependence, especially above 50 K. The magnetic susceptibility below 50 K was fitted to the modified Curie-Weiss law (**Equation 1**)

$$\chi = \chi_0 + \frac{C}{T - \theta_{CW}} \quad \textbf{(Equation 1)}$$

where $\theta_{cw}$ is Curie temperature, $\chi_0$ is the temperature independent magnetic susceptibility, and $C$ is Curie constant. The data was well described by Curie-Weiss model down to 2 K, yielding $\theta_{cw}$ of -0.01(1) K and the effective magnetic moment per Ir ion, $\mu_{eff}$, of 0.494(2) $\mu_B$. The temperature-independent magnetic susceptibility $\chi_0$ is equal to 8.9(1) × 10$^{-4}$ emu/mol, which is larger than that of the reported orthorhombic SrIrO$_3$.

**Figure 5b** shows field dependent magnetization of tetragonal SrIrO$_3$. At 300 K, 100 K, and 1.8 K, a clear diamagnetic contribution was observed, becoming significant at high magnetic fields. However, at 10 K, the field-dependent magnetization exhibited unusual behavior, indicating that a standard Curie-Weiss model is insufficient to describe the magnetization behavior of this system. Considering that tetragonal SrIrO$_3$ behaves as a metal above 50 K (see our electrical resistivity data below), the magnetic susceptibility of non-localized conduction electrons should be taken into account. Then another possible interpretation of the magnetization data arises from itinerant magnetism due to intense electron–electron correlations. Consequently, a monotonically increasing temperature dependent magnetization is not always expected in various fields. This can explain the weak magnetization observed and the small fitted effective moment. Note that different contributions in magnetization are not mutually exclusive. Below 50 K, the localization of electrons was enhanced and a possible Curie-Weiss paramagnetism arises from the local moment. The more accurate magnetic structure and overall spin dynamics should be determined through further experiments.

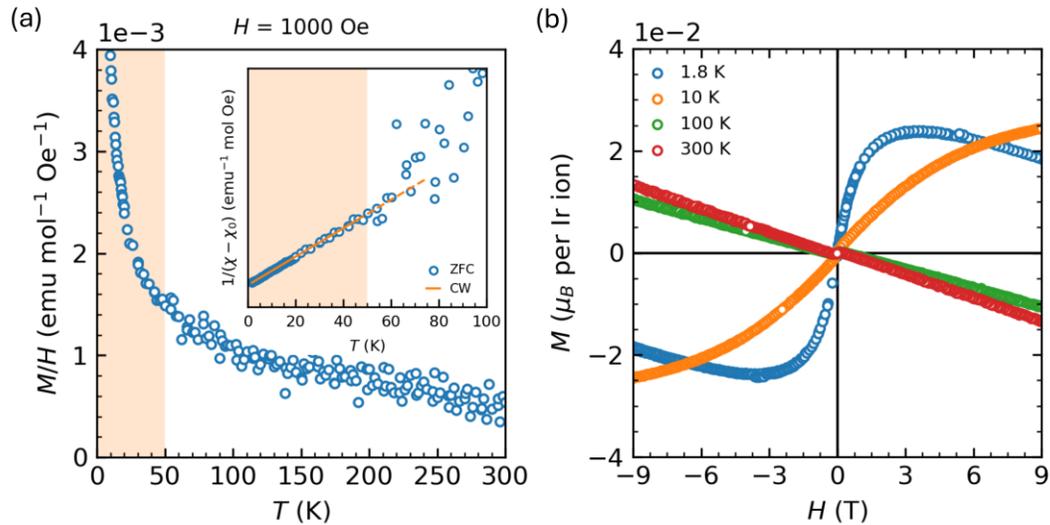

**Figure 5** Magnetization of tetragonal SrIrO$_3$ synthesized under high pressure. **(a)** Magnetic susceptibility measured under ZFC mode at 1000 Oe. Inset, ZFC data fitted by Curie-Weiss model. **(b)** Field-dependent magnetization at 1.8 K, 10 K, 100 K and 300 K.

**Electrical Resistivity: Metal with a Low Temperature Upturn. Figure 6** illustrates the temperature-dependent electrical resistivity of tetragonal SrIrO$_3$ from 2–300 K, noting a room temperature resistivity of 0.038(1) Ω cm. The resistivity data shows metallic behavior with an upturn at 54(2) K, a feature also reported in bulk orthorhombic SrIrO$_3$[22] and monoclinic SrIrO$_3$ thin film systems[15,16,18], typically associated with a metal-insulator transition. Note that in the whole temperature range studied, the electrical resistivity only changed from 0.038 Ω cm to around 0.033 Ω cm, with approximately 13% at maximum. Suggested by magnetization, a possible explanation for the resistivity upturn involves weak electron localization, attributable to intense electron-electron correlations, and possible structural defects in the polycrystalline sample.[30] In iridates, Ir 5$d$ orbitals are highly extended in space and overlap with oxygen $p$ orbitals, which induces interlayer hopping. Different from Sr$_2$IrO$_4$ and Sr$_3$Ir$_2$O$_7$ with two-dimensional [IrO$_6$] interlayer hopping, SrIrO$_3$ with the enhanced interlayer hopping induced by the [IrO$_6$] three-dimensionality becomes, consequently, a metal at room temperature.[10,31] From the electronic state point of view, the bands that cross through the Fermi surface are half-filled and largely split. At low temperatures, electron localization is enhanced, and then the gap opens, contributing to the resistivity upturn. **Figure S6** provides our measurements in the different sequence of cooling and warming, confirming the low-temperature resistivity upturn.

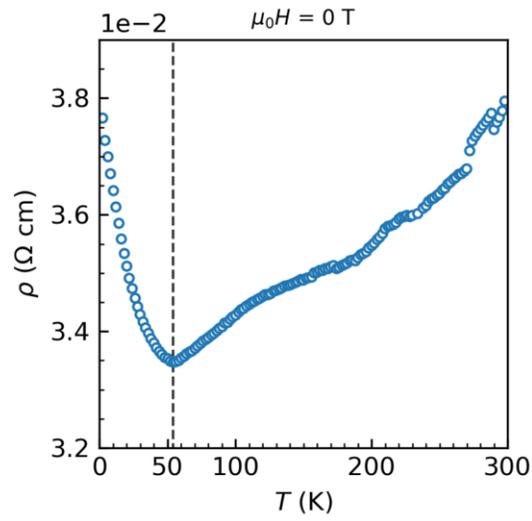

**Figure 6** Temperature-dependent electrical resistivity of tetragonal SrIrO$_3$ at $\mu_0H = 0$ T. An upturn was observed at ~54 K.

## Conclusion

In conclusion, we reported the new tetragonal perovskite SrIrO$_3$ (*P*4/*mmm*, *a* = 3.9362(9) Å, *c* = 7.880(3) Å) synthesized at 6 GPa and 1400 °C. The tetragonal crystal structure was determined based on single crystal and powder XRD results with exclusion of pseudo cubic symmetry. Weak fractional peaks in the H and K dimensions suggest possible structure modulation by oxygen defects. Magnetization study indicated weak paramagnetic behavior down to 2 K, with a small fitted effective magnetic moment of 0.494(2) $\mu_B$, resulting from itinerate electrons. Moreover, electrical resistivity study demonstrated metallic property with an upturn at 54(2) K, attributed to weak electron localization and structural defects. This tetragonal SrIrO$_3$ provides another example that how high pressure can induce novel physical properties in quantum materials.

## Acknowledgments

The work at Michigan State University was supported by U.S. DOE-BES under Contract DE-SC0023648. The work at Brown University was supported by the U.S. Department of Energy, Office of Science, Office of Basic Energy Sciences, under Award Number DE-SC0021265. This work is based on research conducted at the Center for High-Energy X-ray Sciences (CHEXS), which is supported by the National Science Foundation (BIO, ENG and MPS Directorates) under award DMR-1829070.

## Supporting Information

Regenerated reciprocal lattice planes of t*P*-SrIrO$_3$, c*P*-SrIrO$_3$, and o*P*-SrIrO$_3$; SEM and EDS analysis of a single crystal sample; zoom-in powder XRD Rietveld refinement of t*P*-SrIrO$_3$ and o*P*-SrIrO$_3$; magnetic susceptibility of t*P*-SrIrO$_3$; electrical resistivity of t*P*-SrIrO$_3$ in different cooling and warming modes; structure parameters from powder XRD Rietveld refinement of m*C*-SrIrO$_3$, m*C*-SrIrO$_3$+SiO$_2$, t*P*-SrIrO$_3$+IrO$_2$, c*P*-SrIrO$_3$+IrO$_2$, and o*P*-SrIrO$_3$+IrO$_2$; crystallographic data and single crystal XRD refinement of t*P*-SrIrO$_3$ and c*P*-SrIrO$_3$; atomic coordinates and equivalent isotropic atomic displacement parameters.

# Supporting Information

## Pseudosymmetry in Tetragonal Perovskite SrIrO$_3$ Synthesized under High Pressure

Haozhe Wang[1], Alberto de la Torre[2], Joseph T. Race[3], David Walker[4], Patrick M. Woodward[3], Kemp W. Plumb[2], Weiwei Xie[1]*

1. Department of Chemistry, Michigan State University, East Lansing, Michigan 48824
2. Department of Physics, Brown University, Providence, Rhode Island 02912
3. Department of Chemistry and Biochemistry, The Ohio State University, Columbus, Ohio 43210
4. Lamont Doherty Earth Observatory, Columbia University, Palisades, New York 10964

* Email: xieweiwe@msu.edu

## Table of Contents

**Figure S1** Regenerated reciprocal lattice planes of t*P*-SrIrO$_3$, c*P*-SrIrO$_3$, and o*P*-SrIrO$_3$.

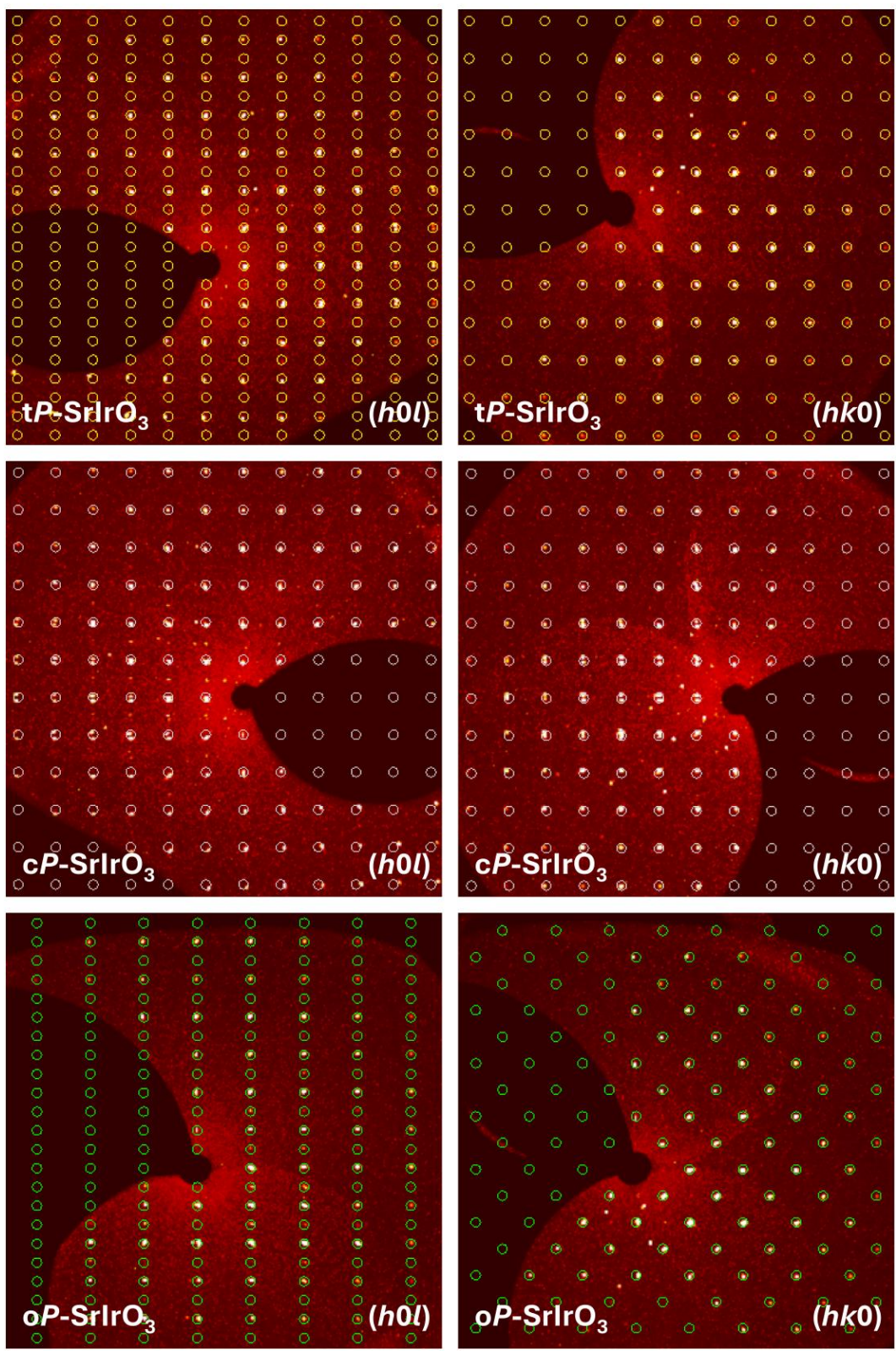

**Figure S2** SEM and EDS analysis of a single crystal sample. The formula obtained was calibrated to Ir and listed in the figure.

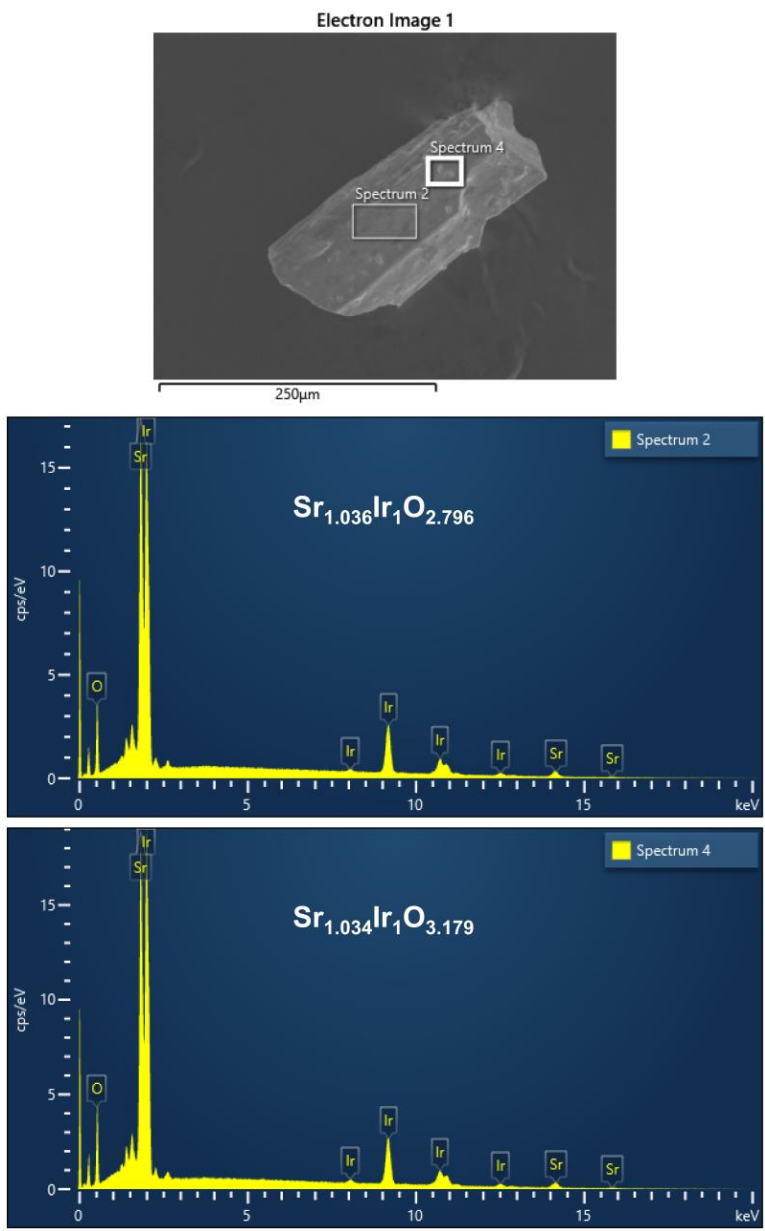

**Figure S3** EDS mapping analysis of a single crystal sample. This confirms the homogenous chemical element distribution of the sample.

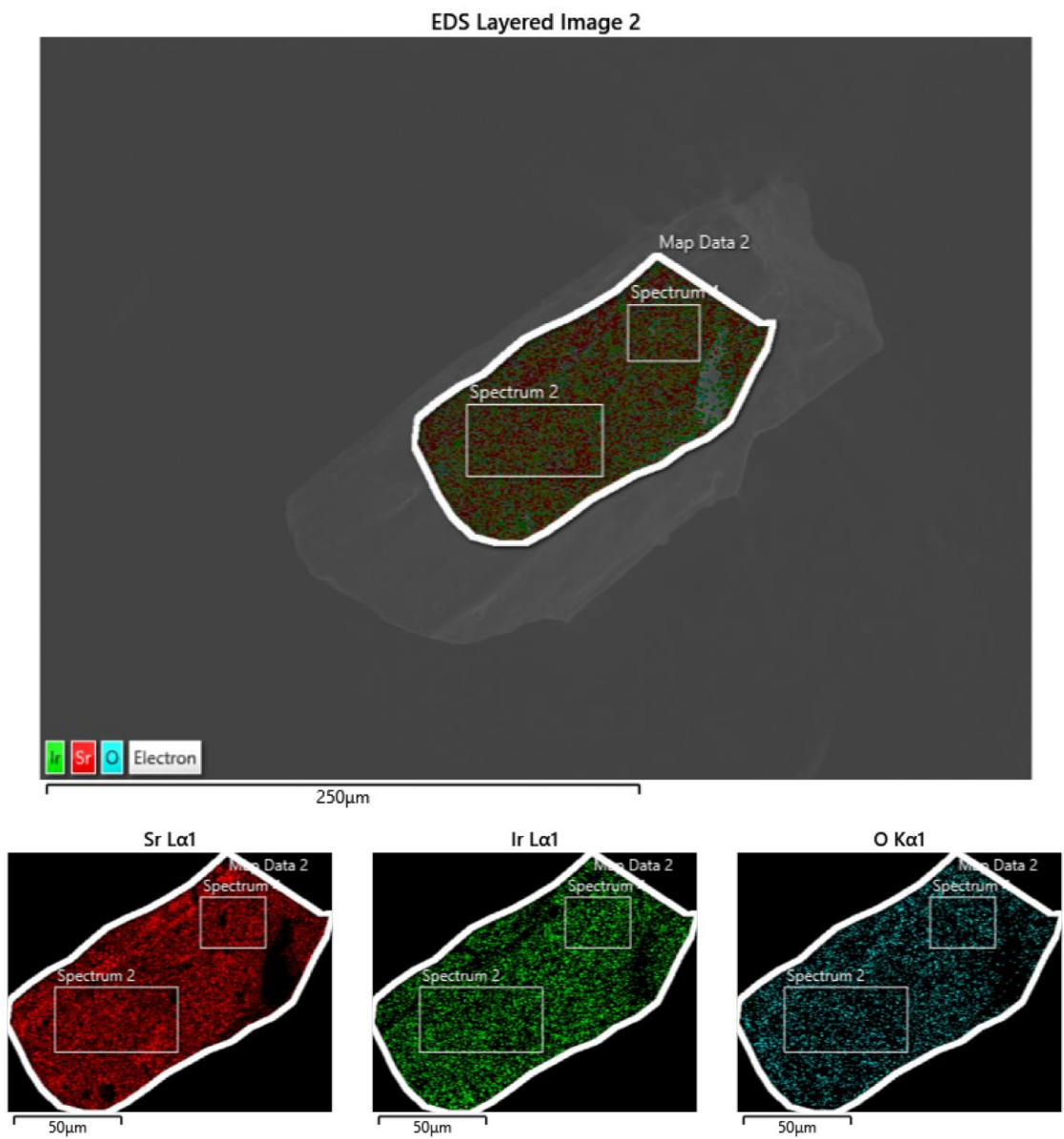

**Figure S4** Zoom-in powder XRD Rietveld refinements of t*P*-SrIrO$_3$ and o*P*-SrIrO$_3$.

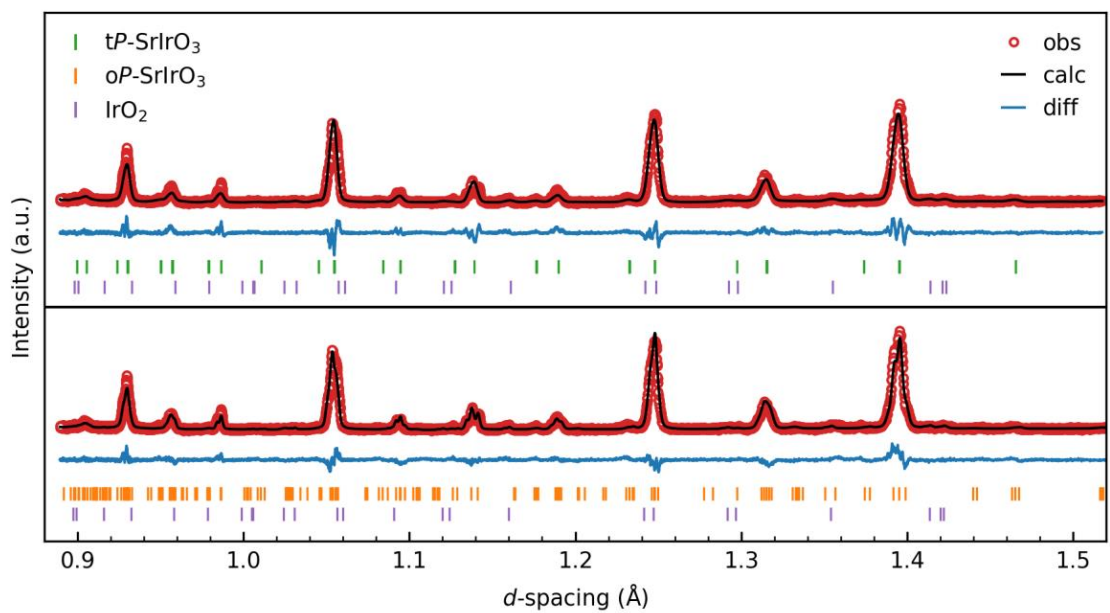

**Figure S5** Temperature-dependent magnetic susceptibility in ZFC and FC modes at 1000 Oe. No significant split between ZFC and FC curves was observed.

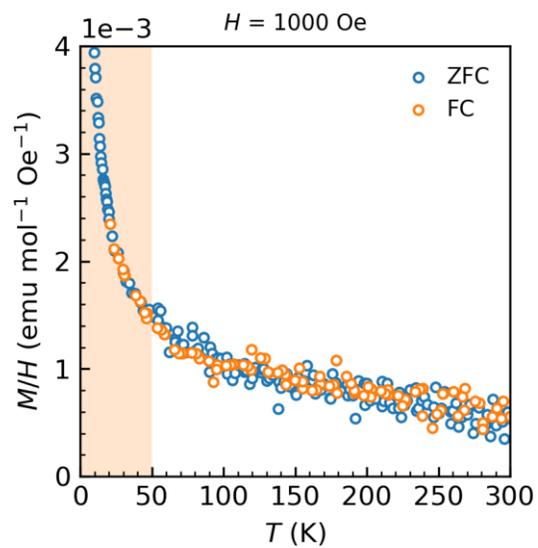

**Figure S6** Temperature-dependent electrical resistivity of t*P*-SrIrO$_3$ in different cooling and warming modes.

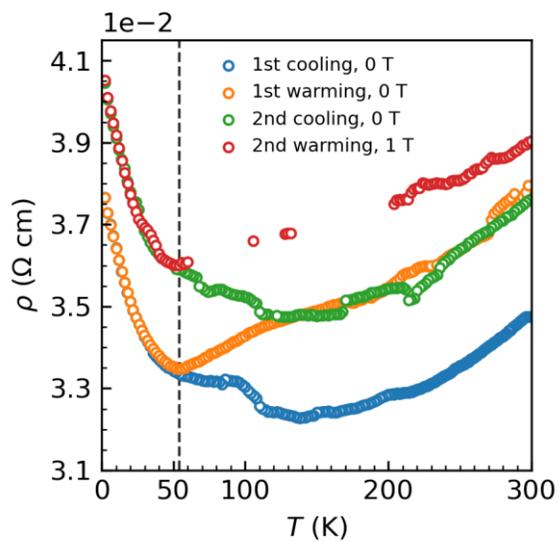

**Table S1** Structure parameters from powder XRD Rietveld refinement of m$C$-SrIrO$_3$.

| Compound | m$C$-SrIrO$_3$ |
|---|---|
| Symmetry | $C2/c$ |
| Unit cell dimensions | $a$ = 5.60047(27) Å |
|  | $b$ = 9.6254(4) Å |
|  | $c$ = 14.1673(8) Å |
|  | $\beta$ = 93.1816(28)° |
| Volume | 762.53(8) Å$^3$ |
| $wR$ | 13.796% |
| Goodness-of-fit | 2.11 |
| Reduced $\chi^2$ | 4.46 |

**Table S2** Atomic coordinates and equivalent isotropic atomic displacement parameters (Å$^2$). ($U_{eq}$ is defined as one third of the trace of the orthogonalized $U_{ij}$ tensor.) $U_{eq}$ values of all the oxygen atoms were constrained to be equal.

| m$C$-SrIrO$_3$ | Wyck. | $x$ | $y$ | $z$ | Occ. | $U_{eq}$ |
|---|---|---|---|---|---|---|
| Ir$_1$ | 4$a$ | 0 | 0 | 0 | 1 | 0.0301(13) |
| Ir$_2$ | 8$f$ | 0.9803(4) | 0.66351(31) | 0.84596(13) | 1 | 0.0347(8) |
| Sr$_1$ | 4$e$ | 0 | 0.9980(9) | 1/4 | 1 | 0.0436(26) |
| Sr$_2$ | 8$f$ | 0.0107(7) | 0.6655(7) | 0.09845(27) | 1 | 0.0382(17) |
| O$_1$ | 4$e$ | 0 | 0.570(4) | 1/4 | 1 | 0.065(5) |
| O$_2$ | 8$f$ | 0.203(5) | 0.2647(25) | 0.2606(16) | 1 | 0.065(5) |
| O$_3$ | 8$f$ | 0.828(4) | 0.3653(26) | 0.0416(15) | 1 | 0.065(5) |
| O$_4$ | 8$f$ | 0.901(5) | 0.157(3) | 0.3995(14) | 1 | 0.065(5) |
| O$_5$ | 8$f$ | 0.300(5) | 0.415(3) | 0.0933(16) | 1 | 0.065(5) |

**Table S3** Structure parameters from powder XRD Rietveld refinement of m$C$-SrIrO$_3$ + SiO$_2$.

| Compound | m$C$-SrIrO$_3$ |
|---|---|
| Symmetry | $C2/c$ |
| Unit cell dimensions | $a$ = 5.60065(26) Å |
| | $b$ = 9.6260(4) Å |
| | $c$ = 14.1681(8) Å |
| | $\beta$ = 93.1818(27)° |
| Volume | 762.65(7) Å$^3$ |
| $wR$ | 12.993% |
| Goodness-of-fit | 1.99 |
| Reduced $\chi^2$ | 3.96 |
| Weight fraction | m$C$-SrIrO$_3$: 0.928 |
| | SiO$_2$: 0.072 |

**Table S4** Atomic coordinates and equivalent isotropic atomic displacement parameters (Å$^2$). ($U_{eq}$ is defined as one third of the trace of the orthogonalized $U_{ij}$ tensor.) $U_{eq}$ values of all the oxygen atoms were constrained to be equal.

| m$C$-SrIrO$_3$ | Wyck. | $x$ | $y$ | $z$ | Occ. | $U_{eq}$ |
|---|---|---|---|---|---|---|
| Ir$_1$ | 4$a$ | 0 | 0 | 0 | 1 | 0.0312(13) |
| Ir$_2$ | 8$f$ | 0.9804(4) | 0.66219(30) | 0.84618(14) | 1 | 0.0360(8) |
| Sr$_1$ | 4$e$ | 0 | 0.0008(9) | 1/4 | 1 | 0.0374(30) |
| Sr$_2$ | 8$f$ | 0.0110(7) | 0.6680(6) | 0.09820(26) | 1 | 0.0317(19) |
| O$_1$ | 4$e$ | 0 | 0.560(4) | 1/4 | 1 | 0.071(5) |
| O$_2$ | 8$f$ | 0.214(4) | 0.2655(24) | 0.2594(15) | 1 | 0.071(5) |
| O$_3$ | 8$f$ | 0.827(4) | 0.3692(25) | 0.0411(14) | 1 | 0.071(5) |
| O$_4$ | 8$f$ | 0.897(5) | 0.145(3) | 0.3950(13) | 1 | 0.071(5) |
| O$_5$ | 8$f$ | 0.283(5) | 0.415(3) | 0.0917(16) | 1 | 0.071(5) |

**Table S5** Crystallographic data and single crystal XRD refinement of t$P$-SrIrO$_3$.

| Chemical Formula | SrIrO$_3$ |
| --- | --- |
| Formula weight | 327.82 g/mol |
| Space Group | $P4/mmm$ |
| Unit cell dimensions | $a$ = 3.9362(9) Å |
|  | $c$ = 7.880(3) Å |
| Volume | 122.09(7) Å$^3$ |
| Density (calculated) | 8.917 g/cm$^3$ |
| Absorption coefficient | 76.005 mm$^{-1}$ |
| $F$(000) | 278 |
| 2$\theta$ range | 5.16 to 79.32° |
| Total Reflections | 1131 |
| Independent reflections | 243 [$R_{int}$ = 0.0438] |
| Refinement method | Full-matrix least-squares on $F^2$ |
| Data / restraints / parameters | 243 / 0 / 18 |
| Final $R$ indices | $R_1$ ($I$>2$\sigma$($I$)) = 0.0596; $wR_2$ ($I$>2$\sigma$($I$)) = 0.1247 |
|  | $R_1$ (all) = 0.0648; $wR_2$ (all) = 0.1289 |
| Largest diff. peak and hole | +14.926 e/Å$^{-3}$ and -12.448 e/Å$^{-3}$ |
| R.M.S. deviation from mean | 1.467 e/Å$^{-3}$ |
| Goodness-of-fit on F$^2$ | 1.122 |

**Table S6** Atomic coordinates and equivalent isotropic atomic displacement parameters (Å$^2$). ($U_{eq}$ is defined as one third of the trace of the orthogonalized $U_{ij}$ tensor.)

| t$P$-SrIrO$_3$ | Wyck. | $x$ | $y$ | $z$ | Occ. | $U_{eq}$ |
| --- | --- | --- | --- | --- | --- | --- |
| **Ir** | 2$g$ | 0 | 0 | 0.25009(13) | 1 | 0.0021(4) |
| **Sr$_1$** | 1$d$ | 1/2 | 1/2 | 1/2 | 1 | 0.0002(6) |
| **Sr$_2$** | 1$c$ | 1/2 | 1/2 | 0 | 1 | 0.053(4) |
| **O$_1$** | 4$i$ | 0 | 1/2 | 0.248(10) | 1 | 0.09(2) |
| **O$_2$** | 1$b$ | 0 | 0 | 1/2 | 1 | 0.11(7) |
| **O$_3$** | 1$a$ | 0 | 0 | 0 | 1 | 0.07(4) |

**Table S7** Crystallographic data and single crystal XRD refinement of c$P$-SrIrO$_3$.

| Chemical Formula | SrIrO$_3$ |
|---|---|
| Formula weight | 327.82 g/mol |
| Space Group | *Pm-3m* |
| Unit cell dimensions | $a$ = 3.9403(6) Å |
| Volume | 61.18(3) Å$^3$ |
| Density (calculated) | 8.898 g/cm$^3$ |
| Absorption coefficient | 75.841 mm$^{-1}$ |
| *F*(000) | 139 |
| $\theta$ range | 5.174 to 39.622° |
| Total Reflections | 510 |
| Independent reflections | 60 [$R_{int}$ = 0.0389] |
| Refinement method | Full-matrix least-squares on $F^2$ |
| Data / restraints / parameters | 60 / 0 / 6 |
| Final *R* indices | $R_1$ ($I>2\sigma(I)$) = 0.0215; $wR_2$ ($I>2\sigma(I)$) = 0.0537 |
| | $R_1$ (all) = 0.0215; $wR_2$ (all) = 0.0537 |
| Largest diff. peak and hole | +2.835 e/Å$^{-3}$ and -5.662 e/Å$^{-3}$ |
| R.M.S. deviation from mean | 0.594 e/Å$^{-3}$ |
| Goodness-of-fit on F$^2$ | 1.329 |

**Table S8** Atomic coordinates and equivalent isotropic atomic displacement parameters (Å$^2$). ($U_{eq}$ is defined as one third of the trace of the orthogonalized $U_{ij}$ tensor.)

| c$P$-SrIrO$_3$ | Wyck. | $x$ | $y$ | $z$ | Occ. | $U_{eq}$ |
|---|---|---|---|---|---|---|
| **Ir** | 1*b* | 1/2 | 1/2 | 1/2 | 1 | 0.0021(3) |
| **Sr** | 1*a* | 0 | 0 | 0 | 1 | 0.0131(5) |
| **O** | 3*c* | 1/2 | 1/2 | 0 | 1 | 0.091(12) |

**Table S9** Structure parameters from powder XRD Rietveld refinement of t$P$-SrIrO$_3$ + IrO$_2$.

| Compound | t$P$-SrIrO$_3$ |
|---|---|
| Symmetry | $P4/mmm$ |
| Unit cell dimensions | $a$ = 3.94447(22) Å |
|  | $c$ = 7.8938(8) Å |
| Volume | 122.818(6) Å$^3$ |
| $wR$ | 16.957% |
| Goodness-of-fit | 2.37 |
| Reduced $\chi^2$ | 5.62 |
| Weight fraction | t$P$-SrIrO$_3$: 0.971 |
|  | IrO$_2$: 0.029 |

**Table S10** Atomic coordinates and equivalent isotropic atomic displacement parameters (Å$^2$). ($U_{eq}$ is defined as one third of the trace of the orthogonalized $U_{ij}$ tensor.)

| t$P$-SrIrO$_3$ | Wyck. | $x$ | $y$ | $z$ | Occ. | $U_{eq}$ |
|---|---|---|---|---|---|---|
| **Ir** | 2$g$ | 0 | 0 | 0.24197(18) | 1 | 0.00034(24) |
| **Sr$_1$** | 1$d$ | 1/2 | 1/2 | 1/2 | 1 | 0.0110 |
| **Sr$_2$** | 1$c$ | 1/2 | 1/2 | 0 | 1 | 0.0278(11) |
| **O$_1$** | 4$i$ | 0 | 1/2 | 0.2824(22) | 1 | 0.0235 |
| **O$_2$** | 1$b$ | 0 | 0 | 1/2 | 1 | 0.0235 |
| **O$_3$** | 1$a$ | 0 | 0 | 0 | 1 | 0.0235 |

**Table S11** Structure parameters from powder XRD Rietveld refinement of c$P$-SrIrO$_3$ + IrO$_2$.

| Compound | c$P$-SrIrO$_3$ |
|---|---|
| Symmetry | $Pm$-$3m$ |
| Unit cell dimensions | $a$ = 3.94478(7) Å |
| Volume | 61.386(3) Å$^3$ |
| $wR$ | 17.524% |
| Goodness-of-fit | 2.45 |
| Reduced $\chi^2$ | 6.00 |
| Weight fraction | c$P$-SrIrO$_3$: 0.974 |
|  | IrO$_2$: 0.026 |

**Table S12** Atomic coordinates and equivalent isotropic atomic displacement parameters (Å$^2$). ($U_{eq}$ is defined as one third of the trace of the orthogonalized $U_{ij}$ tensor.)

| c$P$-SrIrO$_3$ | Wyck. | $x$ | $y$ | $z$ | Occ. | $U_{eq}$ |
|---|---|---|---|---|---|---|
| Ir | 1$a$ | 0 | 0 | 0 | 1 | 0.00153(24) |
| Sr | 1$b$ | 1/2 | 1/2 | 1/2 | 1 | 0.0162(5) |
| O | 3$d$ | 1/2 | 0 | 0 | 1 | 0.0191(21) |

**Table S13** Structure parameters from powder XRD Rietveld refinement of o$P$-SrIrO$_3$ + IrO$_2$.

| Compound | o$P$-SrIrO$_3$ |
| --- | --- |
| Symmetry | *Pnma* |
| Unit cell dimensions | $a$ = 5.59521(15) Å |
| | $b$ = 7.89005(25) Å |
| | $c$ = 5.56572(16) Å |
| Volume | 245.707(9) Å$^3$ |
| $wR$ | 15.296% |
| Goodness-of-fit | 2.14 |
| Reduced $\chi^2$ | 4.58 |
| Weight fraction | o$P$-SrIrO$_3$: 0.974 |
| | IrO$_2$: 0.026 |

**Table S14** Atomic coordinates and equivalent isotropic atomic displacement parameters (Å$^2$). ($U_{eq}$ is defined as one third of the trace of the orthogonalized $U_{ij}$ tensor.)

| o$P$-SrIrO$_3$ | Wyck. | $x$ | $y$ | $z$ | Occ. | $U_{eq}$ |
| --- | --- | --- | --- | --- | --- | --- |
| **Ir** | 4$b$ | 1/2 | 0 | 0 | 1 | 0.00144(22) |
| **Sr** | 4$c$ | 0.5251(5) | 1/4 | 0.4838(7) | 1 | 0.0081(5) |
| **O$_1$** | 4$c$ | 0.510(4) | 1/4 | 0.995(11) | 1 | 0.0293 |
| **O$_2$** | 8$d$ | 0.273(4) | 0.016(4) | 0.729(8) | 1 | 0.0293 |